\documentclass{article}
\usepackage{jheppub}

\usepackage[utf8]{inputenc}
\usepackage[english]{babel}
\usepackage{amssymb}
\usepackage{mathtools}
\usepackage{graphicx}
\usepackage{float}
\usepackage{dsfont} 
\usepackage{cancel}
\usepackage{enumitem}
\usepackage{simplewick}
\usepackage{bbm}
\usepackage{braket}
\usepackage{subcaption}
\usepackage{linearA}

\usepackage{footmisc}
\usepackage{epsf,slashed,soul}
\usepackage{pifont}
\usepackage{bbold}
\usepackage{dsfont}
\usepackage{tikz}
\usepackage{verbatim}
\usepackage{placeins}

\hypersetup{
   colorlinks,
   linkcolor={red!50!black},
    citecolor={blue!50!black},
    urlcolor={blue!80!black}
}

\DeclareSymbolFont{usualmathcal}{OMS}{cmsy}{m}{n}
\DeclareSymbolFontAlphabet{\mathcal}{usualmathcal}

\newcommand{\bea}{\begin{eqnarray}}
\newcommand{\eea}{\end{eqnarray}}

\newcommand{\abs}[1]{|#1|}

\newcommand{\be}{\begin{equation}}
\newcommand{\ee}{\end{equation}}
\newcommand{\Op}{\mathcal{O}}
\newcommand{\lv}{\mathcal{L}}

\title{ Emergence of Krylov complexity through quantum walks: An exploration of the quantum origins of complexity}
\author[a]{Dimitrios Patramanis}
\author[b]{and Watse Sybesma}
\emailAdd{patramanisdimitrios@gmail.com}
\emailAdd{watse.sybesma@su.se}
\affiliation[a]{Kavli Institute for Theoretical Sciences, University of Chinese Academy of Sciences, Beijing 100190, China}
\affiliation[b]{Nordita, KTH Royal Institute of Technology and Stockholm University,
Hannes Alfvéns väg 12, 106 91 Stockholm, Sweden}

\abstract{
In this work we study the relationship between quantum random walks on graphs and Krylov/spread complexity. We show that the latter's definition naturally emerges through a canonical method of reducing a graph to a chain, on which we can identify the usual Krylov structure. We use this identification to construct families of graphs corresponding to special classes of systems with known complexity features and conversely, to compute Krylov complexity for graphs of physical interest. The two main outcomes are the analytic computation of the Lanczos coefficients for the SYK model for an arbitrary number $q$ of interacting fermions and the complete characterization of Krylov complexity for the hypercube graph in any number of dimensions. The latter serves as the starting point for an in-depth comparison between Krylov and circuit complexities as they purportedly arise in the context of black holes. We find that while under certain conditions Krylov complexity follows the growth and saturation pattern ascribed to such systems, the timescale at which saturation happens can generally be shorter than what is predicted by random unitary circuits, due to the effects of quantum speed-ups commonly occurring when comparing quantum and classical random walks.

}

\begin{document}

\maketitle

\section{Introduction}
As we have just concluded the international year of quantum science and technology, we are called to reflect upon the changes that the quantum revolution has brought across all areas of human activity. Arguably, one of the most widely influential ideas that emerged as part of this revolution is the notion of quantum mechanical uncertainty, bringing forth a transition from a deterministic to a probabilistic description of the microcosmos. A particle could be ``here" or ``there", its wavefunction spreading across space and instilling an inherent quantum randomness in our understanding of its behaviour. To say that this change in our way of thinking about the physical world induced a paradigm shift is a gross understatement.

Strangely enough though, the theory of stochastic processes, tasked with the description of randomness in nature, appears to have evolved simultaneously with quantum mechanics, but almost completely independently. In fact, while the foundations were set in the early 20th century by the pioneering works of Markov, Einstein, Smoluchowski and others, several key advances did not come until the middle of the same century (for example Monte Carlo methods and the stochastic integral formulation due to Itô and Stratonovich), that is after quantum mechanics was already a full-fledged theory.

What is the difference then between these two types of randomness and how do they interact with one another? This is the question that served as the source of a myriad confusions and ultimately led to the present work in which we study the notion of complexity for quantum random walks, a topic precisely at the crux of these considerations. Quantum computational complexity is a quantity that has grown to be one of the main objects of interest in the study of complex physical systems. While early works such as \cite{Watrous:2008any,Nielsen1,Nielsen2,Nielsen3} set the stage, it was the works of Susskind and collaborators \cite{Susskind:2014moa,Susskind:2014jwa,Brown:2016wib,Brown:2015bva,Brown:2015lvg,Susskind:2018pmk} that brought complexity to the center of attention, as a candidate to describe the holographic behaviour of wormholes and black holes. This soon made complexity one of the staple points of discourse in the high energy/quantum gravity community. Since then, there has been a proliferation of works computing complexity in a variety of settings. 

Initially, these were mainly concerned with the so-called Nielsen complexity (indicatively see \cite{Chapman:2016hwi,Chapman:2017rqy,Jefferson:2017sdb,Chapman:2018hou} or \cite{Chapman:2021jbh} for a review), but in recent years the field has been dominated by a new measure called Krylov complexity, which will also be one of our main tools in this work. 
Krylov or K-complexity was first introduced in \cite{Parker:2018yvk} as a way to quantify operator growth and it was extended to the description of the complexity of states under the term ``spread complexity" in \cite{Balasubramanian:2022tpr}. Due to its universal features and tractability, this measure saw an explosion of activity with works ranging from holographic considerations \cite{Rabinovici:2023yex,Caputa:2024sux,Das:2024tnw,Heller:2024ldz,Balasubramanian:2024lqk,Heller:2025ddj}, to explorations of quantum chaos and integrability \cite{Dymarsky:2019elm,Dymarsky:2021bjq,Avdoshkin:2022xuw,Camargo:2022rnt,Balasubramanian:2023kwd,Camargo:2023eev,Camargo:2024deu,Chen:2024imd}, as well as properties of condensed matter systems \cite{Caputa:2022eye,Caputa:2022yju,Caputa:2025ucl,Bento:2023bjn,Chakrabarti:2025hsb}. Some advances of central importance to our work here relate to the general symmetry structure underlying the concept of Krylov complexity, which was investigated in the authors' previous works \cite{Caputa:2021sib, Patramanis:2021lkx,Patramanis:2023cwz,Caputa:2023vyr} and several others \cite{Muck:2022xfc,Hornedal:2022pkc,Hornedal:2023xpa,Caputa:2025mii,Caputa:2025ozd}.

Now let us shift our attention to quantum walks which, much like their classical counterpart, have found a wide range of applications both on the theoretical side, as tools for the study of quantum algorithms and universal models of quantum computation \cite{Childs:2008vkn,Lovett:2010uil,Underwood:2010ebd}, and on more practical grounds as models of quantum transport \cite{Karamlou:2021dqa,xu2021quantum} even for processes like photosynthesis \cite{Mohseni:2008mgz} and DNA assembly \cite{Varsamis:2023msp}.
These major results and a general overview on the topic of quantum walks and their importance can be found in a number of reviews \cite{Kempe:2003vul,Venegas-Andraca:2012zkr, Biamonte:2019wdj, Qiang:2024hyh}. Put simply, quantum walks can efficiently describe a number of quantum mechanical phenomena and are thus useful tools both for their theoretical and experimental investigation. As a result of this ubiquity of random processes in nature and especially in the context of complex systems, we believe that the study of complexity is a natural and potentially fruitful direction to explore. While there are many versions of quantum walks that one can define, here we will exclusively concern ourselves with continuous time quantum walks (CTQWs).

In this work we take a two-way approach in uncovering the relation between these two big areas of research. First, we will illustrate that the definitions of a quantum walk and that of Krylov complexity are intertwined and prove that for any graph on which we can define a CTQW, there secretly lies a so-called Krylov chain. We will then use this to relate different classes of systems with distinctive behaviour in terms of their complexity to certain families of graphs. Subsequently, we will do the reverse and use the implicit Krylov structure to study certain CTQWs on graphs of importance. This topic was previously partially investigated in \cite{Jeevanesan:2023ogo}, where the author numerically computes the Krylov complexity for a variety of different graphs. Also in \cite{Sahu:2024kho}, the author discusses Krylov complexity in the context of discrete quantum walks. While these results serve as a useful guide, here we take a more formal and encompassing approach, mostly concerning the fundamental aspects of the interplay between quantum walks and Krylov complexity. 

Finally, let us give a brief account of the contents of this work and its organization. Rather than extensively reviewing the two main topics involved, we have made the choice to ``rediscover" Krylov complexity from the point of view of CTQWs. We believe this approach will be useful to readers who might only be familiar with the latter, while hopefully Krylov complexity experts will find the mental exercise of this rediscovery refreshing. Therefore, we will begin section \ref{QW and krylov} by studying how the dynamics of a quantum walk on a graph can be reduced to that of a chain. Initially, we will discuss this idea for a specific graph whose structure is convenient for that purpose and which will serve as our guide for generalization to arbitrary graphs. Naturally, once we have derived the main concepts and structures surrounding Krylov complexity, we devote a subsection to providing the general context outside of quantum walks, thus highlighting how our approach fits into this broader area of research. We will then comment on the importance of the initial state for our considerations and conclude with the construction of several types of graphs related to systems with notable complexity properties. 

In section \ref{examples section} we will investigate different graphs that arise in the study of physical systems. Most notable among them are the graph that describes the dynamics of the so-called SYK model and the graph of the hypercube in arbitrary dimensions. The latter is one of the graphs favoured by the quantum information community for its helpful properties that we will see in detail below and it has already been the subject of several works on quantum walks (see e.g. \cite{Moore:2001zik,Christandl:2004zdd,Singh:2020bjm}). We will see that the methods that we develop here can serve as a useful tool for the characterization of physical processes that are a priori not related to a quantum walk protocol.

In section \ref{comparison section} we perform an in-depth comparison of the Krylov and circuit complexities arising from quantum and classical random walks on a given graph respectively. Our results in that section are presented as part of the discourse concerning the dynamics of black holes, though they are applicable more generally. We conclude with section \ref{discussion} where we summarize our main findings and discuss their implications.

\section{Quantum walks and Krylov complexity} \label{QW and krylov}
The main idea behind the definition of a continuous time quantum walk (CTQW) on a graph is described in \cite{Farhi:1997jm}, in analogy with the classical version of the problem. Namely, the authors suggest the use of the adjacency matrix of a given graph as the Hamiltonian of quantum evolution. The adjacency matrix encodes the connectivity of a graph. Its elements are 1 if two vertices are connected by an edge and 0 otherwise. Classically, we can derive the transition matrix $W$ directly from the adjacency matrix by imposing the conservation of probabilities, which gives the master equation for their evolution
\be
\overline{p}(t)=e^{\mathcal{W}t}\overline{p}(0)~,
\ee
where $\mathcal{W}$ is the transition matrix $W$ where every column is separately normalized.
The vector $\overline{p}(t)$ tracks the probability of finding the random walker at a given vertex.
If we are to use $W$ for the evolution of the wave functions according to the time evolution equation
\be
\ket{\psi(t)}=e^{-iWt}\ket{\psi(0)}~,
\ee
then $W$ is not subject to  any normalization conditions. 
This emphasizes that the form of the time evolution (unitary or not) is at the heart of why the behaviour of a classical walker is so radically different from a quantum one, as seen for example in \cite{Childs:2001xhf}.
\subsection{Reduction of graph dynamics to a chain } \label{G_4 section}

\begin{figure}[ht]
    \centering
    \includegraphics[width=0.4\linewidth]{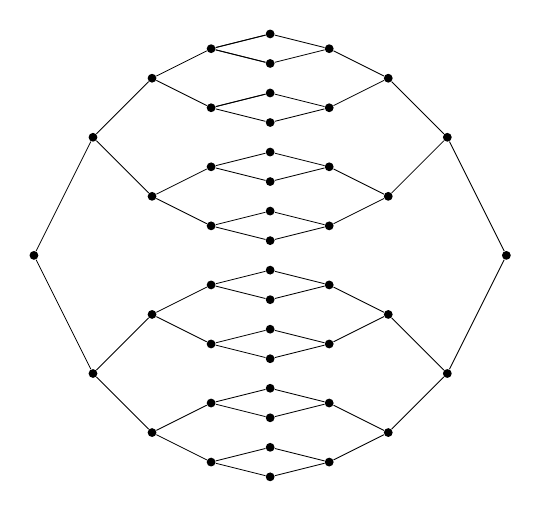}
    \caption{The graph G$_4$, which is the outcome of the fusion of two complete binary trees of order 4 at the leaves. The vertical organization of vertices illustrates the columns mentioned the main text. }
    \label{G4}
\end{figure}

Instead of starting with the definition of Krylov complexity, we will follow the example of a CTQW on the graph of figure \ref{G4}, which was first investigated in \cite{Childs:2001xhf}. What we will find out, is that a notion of complexity naturally emerges as the distance from the origin on the graph. We will then generalize this notion to arbitrary graphs and identify it as the usual Krylov complexity.

The example concerns the CTQW of a single particle on the graph of figure \ref{G4}, assuming the walker starts at the leftmost vertex. In order to analyze the dynamics, the authors of \cite{Childs:2001xhf} show that they can reduce this problem to a quantum walk on the line. Their method relies on the definition of the so-called \textit{column} states 
\be
\ket{j}=\frac{1}{\sqrt{V_j}}\sum_{\alpha \in j}\ket{\alpha}~,
\ee
where $V_j$ is the number of vertices within a column $j$ and $\ket{\alpha_{j}}$ are the states corresponding to these vertices. The summation over $\alpha \in j$ implies summing over every row entry in a particular column. It is implicitly assumed that these states are properly orthonormalized and thus satisfy $\braket{\alpha|\beta}=\delta_{\alpha \beta}$. 
This definition of the column states allows us to represent the collection of vertices within a column as a single vertex. The fact that from each column only stem edges connecting it to the previous and the next ones, guarantees that the action of the Hamiltonian (adjacency matrix) on the column states has the form 
\be
H\ket{j}=a_j\ket{j}+b_{j+1}\ket{j+1} +b_{j}\ket{j-1}~,
\ee
where we have suggestively denoted the transition amplitudes from one column to the next as $a_j$ and $b_j$ since these will turn out to be the so-called Lanczos coefficients. In \cite{Childs:2001xhf} the authors make use of the simple structure of the Hamiltonian\footnote{In this case the authors use the Laplacian of the graph as the Hamiltonian, rather than the adjacency matrix, in order to compare with the classical version of the walk. Since we are not bounded by this requirement, in this work we will always use the adjacency matrix.}
\be
H_{lm}=\left \{  \begin{array}{ll}
                  -\gamma,&\quad m\neq l, \text{ $l$ and $m$ connected by an edge}\\
                  0, &\quad m\neq l, \text{ $l$ and $m$ \textit{not} connected  by an edge}\\
                  k \gamma, &\quad \text{if $m=l$, where $k$ is the degree of the vertex}
                \end{array}
              \right. ~,
\ee
with the degree of a vertex defined as the number of edges attached to it. Using the column states, the only non-zero contributions become
\begin{align}
    &\bra{j}H\ket{j\pm1}=-\sqrt{2} \gamma\\
    &\bra{j}H\ket{j}= \left \{  \begin{array}{ll}
                  2\gamma, \quad j=0,2n\\
                  3 \gamma, \quad \text{otherwise}
                \end{array}
              \right. ~.
\end{align}
Here $\gamma$ is a constant quantifying the transition rate between adjacent vertices. From the above, it immediately follows that  $a_j=2\gamma$ except for the initial and final vertices for which it is $a_0=a_{2n}=3\gamma$, and $b_j=-\sqrt{2}\gamma$.

Having completed this analysis, the authors present a formula for the amplitude to travel between two vertices on the chain as a function of time 
\be
\bra{m}e^{-iHt}\ket{l}=e^{-3i\gamma  t}i^{m-l}J_{m-l}(2\sqrt{2}\gamma t)\,,
\ee
with $J_{m-l}$, a Bessel function of order $m-l$.

Here we will argue that the average distance on the chain can serve as a measure of complexity. Intuitively, if we think about the quantum walk as the visualization of a task/algorithm, how fast or how slowly the walker ``percolates" through the graph can serve as an indication of its efficiency. It is then natural to track the average distance from the origin as a measure that quantifies the complexity of this task. While this particular quantity has not been the focal point of research on quantum algorithms as seen through the lens of quantum walks, it is very much in the spirit of works studying notions such as perfect state transfer, the differences of hitting times between classical and quantum walks etc. In the following subsections we will formalize this intuition into a more rigorous language and show how it fits into the broader framework of Krylov complexity.

\subsection{A general reduction scheme}
With the above example as a prototype, we can generalize the method of reducing any graph to a chain using the following prescription. 
\begin{enumerate}
    \item Choose your favourite vertex (or vertices) on a graph as an initial state $\ket{0}$.
    \item Identify its neighborhood (i.e. nearest neighbor vertices) and define the state 
    \begin{equation*}
        \ket{1}=\frac{1}{\sqrt{\abs{V_1}}}\sum_{\alpha_1 \in V_1}\ket{\alpha_1}~,
    \end{equation*}
where $V_1$ denotes the set of vertices in the neighborhood of the initial vertex and $\ket{\alpha_1}$ the states belonging in that set. We use $\abs{V_1}$ to denote the number of elements (cardinality) of $V_1$.
    \item Identify the neighborhood of each of the vertices in $V_1$ excluding vertices that already belong to a previous state. This constitutes the set $V_2$. Subsequently, define the state
    \begin{equation*}
        \ket{2}=\frac{1}{\sqrt{\abs{V_2}}}\sum_{\alpha_2 \in V_2}\ket{\alpha_2}~.
    \end{equation*}
    \item Repeat this process until you run out of vertices. In the case of an infinite graph, repeat this infinite times. 
\end{enumerate}

From this point forward, we shall call this the \textit{neighborhood partition}\footnote{We use the term neighborhood in the graph theoretical sense, in which the neighborhood of a vertex is the set of its adjacent vertices and the neighborhood of a set of vertices is the union of their individual neighborhoods. While the term layer might seem appropriate given its wide use in the context of networks, we purposefully refrain from using it to avoid confusion with the layered drawing of a graph, using the so-called Sugiyama method \cite{Sugiyama81}.} of a graph, abandoning the column terminology. While columns are easy to visualize for most graphs, we find that thinking in terms of neighborhoods makes it explicit that such a partition of a graph is always possible and the terminology better conforms to the mathematical language of graph theory. It also conveniently allows us to switch our notation from $\ket{j}$ to $\ket{n}$ in order to match with the usual notation of the Krylov basis in the literature. The definition of the neighborhood states is then 
\be \label{neighborhood states}
\ket{n}=\frac{1}{\sqrt{\abs{V_n}}}\sum_{\alpha_n \in V_n} \ket{\alpha_n}~.
\ee
We can readily check using the properties of the states $\ket{\alpha_n}$, that the neighborhood states are also orthonormalized and therefore satisfy by construction $\braket{n|m}=\delta_{nm}$. It should also be clear at this point that what the definition of the neighborhood states achieves is the tridiagonalization of the Hamiltonian (due to its double role as adjacency matrix), since
\be \label{neighborhood Hamiltonian}
H\ket{n}=a_n\ket{n} +b_{n+1}\ket{n+1} +b_n\ket{n-1}~,
\ee
implies that the matrix representation of the Hamiltonian in this basis is of the form
\begin{equation}
      H=
     \begin{pmatrix}
a_1 & b_1 & 0 & \cdots & 0 \\
b_1 & a_2 & b_2 & \ddots & \vdots \\
0 & b_2 & a_3 & \ddots & 0 \\
\vdots & \ddots & \ddots & \ddots & b_{n-1} \\
0 & \cdots & 0 & b_{n-1} & a_n
\end{pmatrix}~.
     \end{equation}
This provides an easy way to compute these coefficients as the matrix elements of the Hamiltonian between adjacent neighborhood states
\be \label{graph lanczos}
b_n= \braket{n-1|H|n}= \frac{1}{\sqrt{\abs{V_n} \abs{V_{n-1}}}}\sum_{\alpha_n,\beta_{n-1}}\bra{\beta_{n-1}}H\ket{\alpha_n}~.
\ee
Explicit knowledge of the graph immediately gives us access to the matrix elements of the Hamiltonian in the original adjacency matrix representation, thus making the computation of the $b_n$ tractable. Similarly one finds 
\be \label{graph an}
a_n= \frac{1}{\abs{V_n}}\sum_{\alpha_n,\beta_n}\bra{\beta_n}H\ket{\alpha_n}~.
\ee
 It is also important to highlight that the sums in the above expressions count the total number of edges between two adjacent neighborhoods in the case of $b_n$ and the total number of edges within a neighborhood (if any) in the case of $a_n$. By denoting the number of edges between the $n$th neighborhood and the next as $E_n$, and the number of edges within a neighborhood as $I_n$ (we arrive at this notation by thinking about external and internal edges) we can then rewrite
 \be
b_n= \frac{E_{n-1}}{\sqrt{\abs{V_n}\abs{V_{n-1}}}},\quad a_n=\frac{I_n}{\abs{V_n}}~.
 \ee
 Note that these expressions can be made to hold true for the case of a weighted graph in which the quantum walk is biased (meaning that the Hamiltonian is not just the adjacency matrix, but a weighted version thereof). However, the neighborhood states have to be redefined such that they form an invariant subspace under the action of the  Hamiltonian (for more details see appendix \ref{graph conditions}). The difference is that the sum terms in (\ref{graph lanczos}) and (\ref{graph an}) do not simply count edges any more, but rather one has to keep track of the individual weights when performing the sum. This already adds a layer of sophistication which is beyond the scope of the present work, however we will later discuss an example where this version of a walk is applicable. Generally, it is conceivable that even for a biased walk with simple enough rules for assigning weights to edges, the computation of these coefficients is tractable.  More importantly though, the expression for the $b_n$ highlights a crucial relation between the graph structure and the dynamics. Namely, notice that that the growth of the coefficients does not solely depend on the growth of the number of vertices in each neighborhood as one might have expected. Instead, what is important is the interplay between the growth of the number of vertices and the connectivity of the graph. 
 
 At this point the reduction of a graph to a chain has been completed. The numbers $a_n$ and $b_n$ contain all the information for the evolution on the chain, which can be seen by expressing the time evolved state in the neighborhood basis
 \be
 \ket{\psi(t)}=e^{-i H t}\ket{\psi(0)}= \sum_{n=0}^N \varphi_n(t)\ket{n}~,
 \ee
 where $\varphi_n$ are the wavefunctions on the chain resulting from this decomposition. Acting with a time derivative on this equation reveals the dependence of the evolution on the $a_n,\,b_n$
 \be \label{chain schrodinger}
-i\frac{d \varphi_n(t)}{dt}=a_n\varphi_n(t) + b_{n+1}\varphi_{n+1}(t) +b_n \varphi_{n-1}(t)~.
 \ee
Given the initial condition $\varphi_0=\bra{0}e^{-iHt}\ket{0}$, this Schrödinger equation can be solved iteratively leading to the set of the wavefunctions $\varphi_n$ and the complete characterization of the problem. The average distance on the chain, which we have asserted can serve as a measure of complexity is given by 
\be
\mathcal{C}_K=\sum_{n=0}^N n\abs{\varphi_n(t)}^2~.
\ee
The readers familiar with the notion of Krylov/spread complexity have probably already realized that our so-called neighborhood basis, is just the Krylov basis, the coefficients $a_n$ and $b_n$ are the Lanczos coefficients and $\mathcal{C}_K$, Krylov/spread complexity itself. This is not a matter of happenstance, but rather a rigorous correspondence which emerges as a result of an implicit graph symmetry, however we will delegate the formal details of this statement to appendix \ref{graph conditions}. For the unfamiliar reader let us quickly summarize these definitions and highlight their relationship with the above. While we hope the following relays the core concepts and ideas, the interested reader may refer to \cite{Parker:2018yvk,Balasubramanian:2022tpr,Nandy:2024evd,Baiguera:2025dkc,Rabinovici:2025otw}, for more details.
\subsection{A quick Krylov/spread complexity primer}
Suppose we wish to study the time evolution of a generic quantum system in some initial state $\ket{\psi_0}$ with a given Hamiltonian $H$. One way to do so is by tracking its evolution through the space spanned by the states produced when acting with powers of the Hamiltonian on the initial state
\be
\{\ket{\psi_n}\}=\{H\ket{\psi_0},H^2\ket{\psi_0},...H^N\ket{\psi_0}\}~,
\ee
defining the so-called Krylov subspace.
One reason why we might be interested in that space specifically is that the Hamiltonian is an inherent feature of the system and thus, in principle, this construction is always available to us. However, decomposing the time evolved state in this particular space is not so straightforward since the set $\{\ket{\psi_n}\}$ is not automatically orthonormalized. It follows that in order to provide a proper basis we need to implement an orthonormalization procedure. This is typically done via the Lanczos algorithm. This is just a different name for the Gram-Schmidt process applied to this particular set of vectors, whose special structure implies that Hamiltonian in the basis we obtain is tridiagonal (for more details on that point see \cite{Viswanath1994TheRM}). Consequently, the output of the Lanczos algorithm is the orthonormal Krylov basis and the Lanczos coefficients $a_n$ and $b_n$ corresponding to the diagonal and off-diagonal elements of the Hamiltonian respectively. 

What this achieves is that the Hamiltonian's action on the Krylov basis is of the form (\ref{neighborhood Hamiltonian}). From this we can derive the equation for the motion for a particle on a chain exactly like (\ref{chain schrodinger}), which however is an a priori completely abstract picture not originating from the reduction of a graph. Finally, we can endow the average distance on the chain with an interpretation as a measure of complexity which effectively counts how many times on average one needs to apply the Hamiltonian to the initial state such that they are able to effectively describe the time evolved state.

We should note that instead of tracking the evolution of a quantum state, we might apply this logic to the evolution of quantum operators. Instead of using the Hamiltonian, we can use the Liouvillian super-operator defined as the commutator with the Hamiltonian $\lv=[H,\cdot]$. While the evolution now takes place in the Hilbert space of operators, rather than the usual Hilbert space of states, the logic is exactly the same. We can tridiagonalize the Liouvillian (with a judicious choice of inner product we can also typically set $a_n=0$) and obtain a Krylov chain picture, same as before. 

An important part of this construction is the introduction of bounds on the rate of spread of quantum states, or of the growth of operators. In terms of the Lanczos coefficients the bound takes the form \cite{Parker:2018yvk}
\be \label{Lanczos linear bound}
b_n\leq \gamma n+ O(1)~,
\ee
or in other words the growth of the Lanczos coefficients as a function of the Krylov index is at most linear. Conjecturally, this bound should be saturated by chaotic systems, though the reverse is not necessarily true. In fact there are already well studied examples of free theories which exhibit such maximal growth in the context of QFT (see for example \cite{Dymarsky:2021bjq,Avdoshkin:2022xuw}).

 The graph perspective offered here can also help us establish certain restrictions on the behaviour of the Lanczos coefficients from graph theoretical arguments, albeit much less stringent than the one given above. Denoting the number of edges between the neighborhoods with $n-1$ and $n$ as $E_{n-1}$, it follows that
\be
E_{n-1}\leq \abs{V_{n}}\abs{V_{n-1}}~,
\ee
where the equality holds in the case that every vertex in a given neighborhood connects to every vertex in the next one. Given that our construction relies on the neighborhood partition of a graph, we can also introduce a lower bound for the number of edges, since every vertex in a given neighborhood must have at least one edge connecting it to the previous one. This implies
\be
E_{n-1}\geq \abs{V_n}~.
\ee
At the level of the Lanczos coefficients the above constraints become
\be \label{graph bound}
 \sqrt{\frac{\abs{V_n}}{\abs{V_{n-1}}}}\leq b_n\leq \sqrt{\abs{V_n} \abs{V_{n-1}}}~.
\ee
In principle $\abs{V_n}$ can be an arbitrary function of $n$ (as long as it returns a natural number), which shows that the upper bound is indeed much more lax than (\ref{Lanczos linear bound}) which arises from physical arguments.

\subsection{The choice of initial state}
Previously we assumed that our initial state corresponds to a single vertex of a given graph. Moreover, we typically picked a ``privileged" vertex from which the rest of the graph seems to naturally ``sprout". While this is indeed the reasonable thing to do for a variety of setups and especially when we think about the random walk on the graph as a visualization of an algorithm, in which case we have the freedom to organize the initial conditions as we see fit, it does not have to be the case in principle. Instead we could choose any vertex, or superposition of vertices, as the initial state on which time evolution will be applied. The algorithm we provided remains largely the same, the only difference being that if we wish to start with a superposition of states, the initial neighborhood will be comprised of all the corresponding vertices. The process of performing the neighborhood partition and defining the neighborhood states remains the same. 

However, starting from different initial states can lead to very different Krylov bases -- different Krylov subspaces -- and Lanczos coefficients. This is not at all surprising, since it is already well known that constructing these quantities is largely dependent on the choice of initial state (for a more detailed discussion and a method of mitigation of the effect of this choice see \cite{Craps:2024suj}). Here we will highlight the effect that the choice of initial state might have by investigating a few simple examples and discussing the arising implications. 

\begin{figure}[ht]
    \centering
    \includegraphics[width=0.25\linewidth]{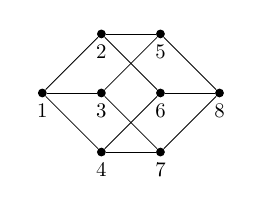}
    \caption{The graph of a cube with its vertices labeled for reference.}
    \label{cube}
\end{figure}

\begin{figure}[ht]
    \centering
    \includegraphics[width=0.2\linewidth]{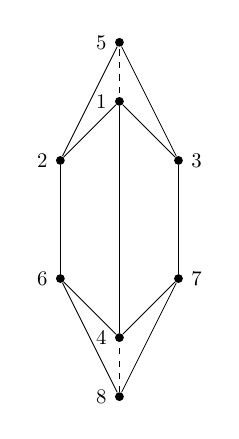}
    \caption{A different version of the quantum walk on the cube, in which the initial state is an equal superposition of vertices 2 and 6. The dashed line expresses vertices 5 and 8 to be connected.}
    \label{shuffled cube}
\end{figure}

We will start by considering the graph of a cube, which apart from providing a good testing ground for our intended purposes, will also serve as a useful warm-up for section \ref{Hypercube section} in which we will treat the hypercube graphs in full generality. The graph of the cube appears in figure \ref{cube}, to which we will refer to for different choices of initial states. 
The standard choice we discussed previously is to choose vertex 1 as our initial state. In that case it is easy to verify that the Krylov space has 4 elements and the Lanczos coefficients are $b_1=\sqrt{3}, \,b_2=2,\, b_3=\sqrt{3}$. Let us now consider a more involved initial state by picking an equal superposition of the states corresponding to vertices 2 and 6. In that case, according to our prescription, we have to rearrange the cube as seen in figure \ref{shuffled cube}.
From this drawing it is evident that there are only three neighborhoods and as a result only three Krylov basis elements. The Lanczos coefficients are 
\begin{align}
  &  a_0= \frac{1}{2}, \quad a_1= \frac{3}{4}, \quad a_2=\frac{1}{2},\\
  & b_1= \sqrt{2}, \quad b_2=\sqrt{2}~.
\end{align}
Clearly, the structure of these two versions of the quantum walk on a cube is quite different. It is difficult to categorize all the different cases, which we can illustrate by considering what happens when we choose a superposition of vertices 1 and 8 as the initial state (this case is a bit harder to draw efficiently so we invite the reader to attempt it for themselves). In that case the structure is yet again different, with only two Krylov basis states and the Lanczos coefficients 
\begin{align}
    & a_0=0,\quad a_1=2,\\
    &b_1=\sqrt{3}~.
\end{align}

It is easy to see that choosing a general superposition as an initial state will have similar effects, which serves to show that this choice can completely change the structure of the problem. It is worth mentioning that if we start with an initial state that is a superposition of all the states corresponding to vertices, the Krylov space is automatically one-dimensional irrespective of any other properties of the graph and however small the coefficient of the superposition might be for any given vertex. The only quantity that remains to be determined is $a_0$ and of course there is not much complexity to speak of. This might be surprising. After all, why should a generic quantum state behave differently from a special choice that we arbitrarily dictated? The answer is that by reducing the dynamics of the graph onto a chain, we forgo many of the fine grained details contained in it. Krylov complexity quantifies the spread of an operator or a state within a particular subspace and so if we start with an operator or state that is already spread out in the whole Hilbert space, there is not much more we can learn about it. In other words, the information contained in the coefficients of the expansion of the original state over the Hilbert space is lost by this reduction. What remains is only the collective information about the distance from the origin. 

With these observations in mind, we move forward, where we will be mostly interested in the simplest case where the initial state corresponds to a single vertex, which is after all one of the most insightful choices.

\subsection{Special classes of Lanczos coefficients and respective graphs} \label{special bn}
There are a few notable classes of Lanczos coefficients that can help us draw distinctions between different kinds of dynamics. These include the case where $b_n\sim\text{constant}$, in which case the complexity will exhibit linear growth, $b_n\sim\sqrt{n}$ which is related to integrable dynamics and $b_n\sim n$ potentially signifying chaos. We will also be examining the case $b_n\sim\sqrt{n(n-1)}$, which along with $b_n\sim\sqrt{n}$, is of additional interest due to its explicit relation with symmetry. In order to construct graphs conforming to these classes there are two variables that we need to balance, namely the number of vertices in a neighborhood $V_n$ (since from this point on we will always refer to the number of elements in a neighborhood and not the set itself we will drop the absolute value for the sake of convenience) and the number of edges between adjacent neighborhoods $E_n$. This can be formalized using (\ref{graph lanczos}) as 
\be \label{Lanczos edges}
b_n=\frac{E_{n-1}}{\sqrt{V_n V_{n-1}}}~.
\ee
We remind the reader that $E_{n}$ is not a completely independent function as the bound on $b_{n}$ (\ref{graph bound}) has to be respected. Moreover, it is important to remember that both $E_{n}$ and $V_{n}$ should be natural numbers (or zero), which is a highly stringent constraint on the form these functions can assume. 

Starting from the simplest case $b_n\sim\text{constant}$, it is easy to find a combination of $E_n, V_n$ producing this result by setting
\begin{align} \label{exp graps}
   & V_n=c_1 q^{c_3 n}~, \\
   &E_n=c_2q^{{c_3(n+1)}}~,
\end{align}
where $c_1,c_2,c_3,q$ are constant natural numbers.
Note that the $c$ constants are technically subject to additional constraints given that the bound (\ref{graph bound}) has to be respected. We will not be very careful about this issue for the families of graphs discussed herein, however one has to keep this in mind when constructing a particular member of a given family. From the above it is easy to compute $b_n=\frac{c_2}{c_1}q^{c_3/2}$. For example we can obtain up to the midpoint of the graph $G_4$ discussed in section \ref{G_4 section} for $c_1=c_{2}=c_{3}=1$, $q=2$. The fact that after the middle of the graph the Lanczos coefficients  behave the same is due to the invariance of the formula (\ref{Lanczos edges}) under the exchange of $V_n$ and $V_{n-1}$. In other words if a graph is growing or shrinking in the same way, the $b_n$ remain the same, which means that given a number of neighborhoods, we can draw several different graphs with regions of growth or shrinkage. In the example above this can be formalized by realizing that 
\be
V_n=\left \{  \begin{array}{ll}
                2^n,&\quad 0\leq n\leq 4\\
                  2^{9-n},&\quad 4<n\leq 9\\
                \end{array}
              \right. ~.
\ee
In other words, we can understand it as the concatenation of two graphs belonging in the same family described by (\ref{exp graps}). This is straightforward to generalize for more complicated ``Frankenstein's Monster" graphs. 

Another point worth stressing is that (\ref{Lanczos edges}) takes into account the total number of edges, but not how they are distributed among the vertices involved. This further increases the possible graphs one can draw for a given set of Lanczos coefficients, while staying in the same family. That said, we will be interested in graphs that exhibit some regularity and so, when possible, we will be dealing with graphs for which the degree of the vertices within a given neighborhood is constant. Apart from being easier to carry out calculations with, we believe these are more likely to describe naturally occurring physical processes, as we will see for example in section \ref{SYK section}. It should be noted however that there is also significant interest in the case of random graphs, albeit for different reasons. One can find an approach similar to the one developed in this work that targets this class of problems in \cite{Balasubramanian:2023qoi}.

\begin{figure}[ht]
    \centering
    \includegraphics[width=0.25\linewidth, angle=270]{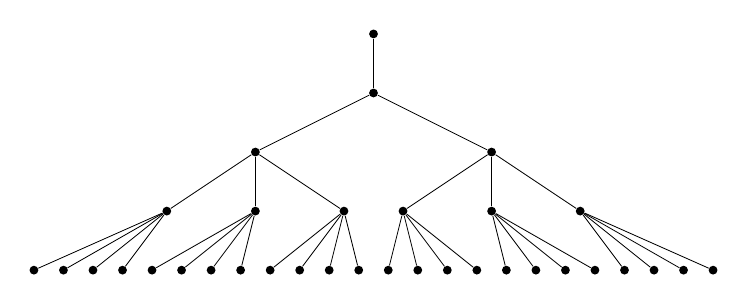}
    \caption{The first 5 neighborhoods, shown from top to bottom, of the factorially growing graph.}
    \label{factorial graph}
\end{figure}

With these clarifications in mind we can move on to the next class of Lanczos coefficients with $b_n\sim \sqrt{n}$. We find that this behaviour arises for a family of graphs described by
\begin{align}
    &V_n=c_1 n!~,\\
    &E_n=c_2 (n+1)!~.
\end{align}
This in turn yields $b_n=\frac{c_2}{c_1}\sqrt{n}$. This is clearly a family of rapidly expanding graphs, but taking as a representative one satisfying the regularity condition given above and for $c_1=c_2=1$, its structure is rather simple and can be seen in figure \ref{factorial graph}.
As mentioned previously this case is of interest because this type of growth for the Lanczos coefficients is associated with integrable dynamics. Moreover, it was shown in \cite{Caputa:2021sib} that Heisenberg-Weyl symmetry inadvertently leads to this result as well. 

\begin{figure}[ht]
    \centering
    \includegraphics[width=0.5\linewidth]{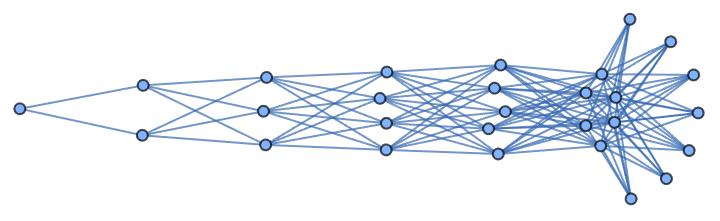}
    \caption{A squid graph of 7 neighborhoods}
    \label{squid graph}
\end{figure}

Next, we will study the case $b_n \sim\sqrt{n(n-1)}$, which is related to dynamics with SL(2,R) symmetry. Note that for large $n$ we obtain $b_n\sim n$, which implies that the complexity growth for such systems is nearly maximal, or in terms of the quantum walk picture, this is the maximal rate at which a walker can penetrate a graph. The solution for the graph structure in this case is exceptionally simple and it takes the form 
\begin{align}
    &V_n=c_1 n\,,\\
    &E_n=c_2 n(n+1)~.
\end{align}
For $c_1=c_2=1$ this leads to the graph of figure \ref{squid graph} which has the characteristic shape of a squid.
Notice that despite this graph not growing very rapidly in terms of the number of vertices within each neighborhood, it exhibits rapid growth in its inter-neighborhood connectivity.                                                             

Having constructed these different families of graphs there are a few remarks that we wish to make. The first concerns uniqueness. While we have already seen that for a given sequence of Lanczos coefficients we can construct many different graphs, we have associated each type of growth for the $b_n$ with a specific family. The question that arises then is whether we can achieve the same growth with different families of graphs. For instance can we achieve $b_n\sim\sqrt{n}$ with something other than the ``factorial" graphs we considered? While a rigorous answer to that question goes beyond the scope of this work (and probably requires the expertise of a graph theorist), we are temped to claim that the answer is no, there are no other families of graphs that achieve the same types of growth we considered. The reason for this claim is the highly constrained nature of the variables involved. As we discussed we require that both $V_n$ and $E_n$ are natural numbers, so there are not that many forms that they can assume while remaining expressible as functions. Among these we have considered polynomial, exponential and factorial functions which to our knowledge comprise a large portion of the available options. Additionally, we have the auxiliary constraint $b_n\leq \gamma n +O(1)$, which further restricts the kinds of graphs that one can consider. 
While this does not constitute a proof of any kind, we have at the very least been unable to find a counterexample, so for the time being we maintain that this is a reasonable claim to make. 

The second remark has to do with the properties of the graphs themselves. We observe that the first two families we presented are planar graphs, whereas the latter two are not. In section \ref{examples section} we will see further examples of important graphs that conform to either one of these two categories and most notably the non-planar graphs having to do with the dynamics of the SYK model and the hypercube. It is possible that this feature might be related to the geometry of the space of states, as we know for example that the Heisenberg-Weyl dynamics we attributed to the factorially expanding graph lead to a flat metric on the space of states (see \cite{Caputa:2021ori}). Similarly the non-planar squid graph is compatible with the negatively curved geometry of the space of states for a system with SL(2,R) symmetry. It would be interesting to investigate whether this observation has merit, whether possibly a flavor relation to large number of color counting is involved \cite{tHooft:1973alw}, or it is merely a mathematical coincidence, but we will leave that for future work. 

The final remark concerns the potential speed-ups one can achieve with quantum walks on graphs belonging to these families. The term speed-up can refer to a variety of quantities, such as the mixing time (time it takes to get ``close" to the steady state solution) and hitting time (time it takes for a walker to have non-zero probability to be found at a given vertex). One of the main reasons for studying quantum walks is in order to find potential ways to speed-up algorithms by improving these times utilizing quantum dynamics. As it was shown for the example of the graph in figure \ref{G4}, concatenating two graphs of the same family such that the neighborhoods grow and then shrink in size can typically lead to speed-ups (in this case the time it takes to traverse the graph experiences an exponential speed-up). The simple explanation behind it, is that a classical walker initially sees many paths forward, but at some point this number caps off. The probability that the walker will then move either forward or backward is the same and so the expected distance saturates at that value. On the contrary a quantum walker's wavefunction will spread throughout the graph uninhibited, hence why the probability of the walker reaching an arbitrary exit vertex is not suppressed. We believe that the majority of the graphs we presented here can lead to speed-ups under similar conditions.

\section{Examples} \label{examples section}
In this section we will investigate several examples of quantum walks on graphs and their respective complexity in more detail. This way we highlight the utility of our approach and provide some concrete insights into the dynamics of different quantum walks.

\subsection{Complete graphs}
A complete graph is one in which every pair of vertices is connected with a unique edge and as we will see it leads to somewhat trivial results when it comes to complexity. The reason is that since our initial state (whether it corresponds to a single vertex or not) connects to all other vertices, there will only be two neighborhoods. Starting with the complete graph with $q$ vertices $K_q$ and choosing a single vertex as our initial state (all vertices in this case are equivalent) we easily obtain from (\ref{graph lanczos})
\be
b_1=\sqrt{q-1}, \quad a_0=0,\quad a_1=q-2~.
\ee
We then find that the complexity scales as 
\be
\mathcal{C}_K\sim \sin^2{\left(\frac{qt}{2}\right)}~.
\ee
This picture corroborates the results of \cite{Jeevanesan:2023ogo}, who found that Krylov complexity as a function of the Hilbert space dimension is minimal for complete graphs. Clearly, the reason why is that complete graphs always have at most two Krylov basis vectors.

\subsection{Circular graphs}
A quantum walk on a circular graph is one of the simplest problems to tackle and for this reason it has been a popular testing ground for a variety of applications (for more details see for example section 3.1 of \cite{Qiang:2024hyh}). We can easily compute the Lanczos coefficients for circular graphs using (\ref{graph lanczos}). For an even number of vertices we obtain
\be
b_n=\left \{  \begin{array}{ll}
                  \sqrt{2}, \quad &n=1 \,\text{and}\, n=N\\
                  1,\quad &\text{otherwise}
                \end{array}
              \right. ,\quad a_n=0~,
\ee
whereas for an odd number of vertices
\be
b_n=\left \{  \begin{array}{ll}
                  \sqrt{2}, \quad &n=1 \\
                  1,\quad &\text{otherwise}
                \end{array}
              \right. ,
        \quad a_n=\left\{  \begin{array}{ll}
                  1, \quad &n=N \\
                  0,\quad &\text{otherwise}
                \end{array}
              \right. ~.
\ee
While a general closed formula for the complexity does not seem to exist for arbitrary graphs, it is easy to compute it explicitly for any given graph using the above. As one would expect, it exhibits an oscillatory pattern like all finite graphs. Unlike the complete graph though, here we can already find an extended Krylov space with a wealth of information about the evolution of the system. In order to get a better sense of the above let us consider the example of a circular graph with 8 vertices and compare it with the complete graph with the same number of vertices. 

\begin{figure}[ht]
   \begin{subfigure}{0.5\textwidth}
       \includegraphics[width=0.9\linewidth]{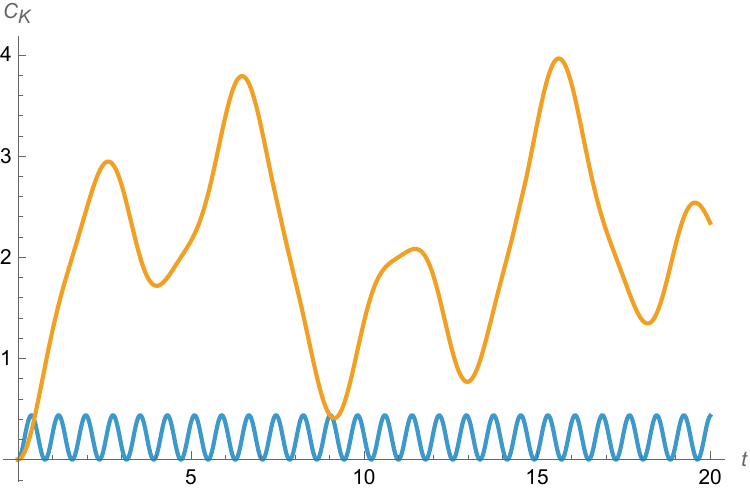}
       \caption{}
\label{cvc}
  \end{subfigure}
\begin{subfigure}{0.5\textwidth}
    \includegraphics[width=0.87\textwidth]{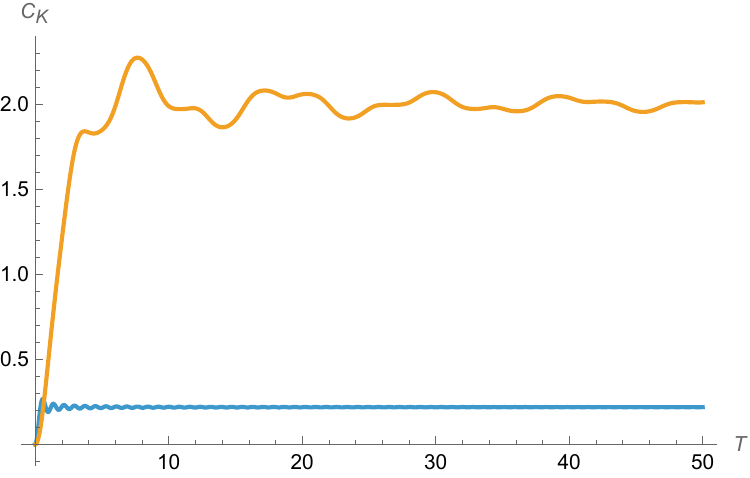}
    \caption{}
\label{cvc av}
\end{subfigure}
\caption{(a) Comparison of the  Krylov complexities for the complete (blue) and circular (orange) graphs of 8 vertices as functions of time. (b) Comparison of the time averaged complexities for the complete (blue) and circular (orange) graphs of 8 vertices.}
\label{circle vs complete}
\end{figure}

For the circular graph we find that the Krylov space is 5-dimensional and the complexity is
\be
\mathcal{C_\text{circle}}=2-\cos (t) \left(\sqrt{2} \sin (t) \sin \left(\sqrt{2} t\right)+2 \cos (t) \cos
   \left(\sqrt{2} t\right)\right)~,
\ee
whereas for the complete graph the Krylov space is only two-dimensional and the complexity is 
\be
\mathcal{C}_\text{complete}=\frac{7}{12}\sin^2{(4t)}~.
\ee
In figure \ref{circle vs complete} 
we give a side by side comparison of the complexities and the time averaged complexities defined as
\be \label{time average complexity}
\overline{\mathcal{C}_K(T)}=\frac{1}{T}\int_0^T\mathcal{C}_K(t)dt~,
\ee
to better illustrate the difference in magnitude. As expected the complete graph exhibits lower maximum complexity and a purely periodic behaviour, whereas the circular graph exhibits a more erratic pattern for its complexity, with a higher saturation value.

\subsection{Operator growth in SYK} \label{SYK section}
The SYK model has been extensively studied as one of the rare examples of systems purportedly exhibiting  chaotic dynamics, while still remaining analytically solvable in certain regimes. It has also attracted a lot of attention as a system with a holographically dual description in terms of low-dimensional gravity \cite{Maldacena:2016hyu}. For these reasons it has been the subject of several works that discuss its Krylov complexity \cite{Rabinovici:2023yex,Ambrosini:2024sre,Bhattacharjee:2022ave,Bhattacharjee:2023uwx,Nandy:2024zcd}. Here we revisit this problem utilizing our newfound graph perspective. In \cite{Bhattacharjee:2022lzy}, the authors also develop a diagrammatic approach to analytically compute the Lanczos coefficients in the limit where the number of interacting fermions $q$ becomes large, however here we give what is to our knowledge the first instance of an analytical expression for any  $q$.

We will use the results of \cite{Roberts:2018mnp}, in which the authors discuss operator growth in the SYK model. In particular, starting from the usual Hamiltonian for N fermions
\be
H=i^{\frac{q}{2}}\sum_{1\leq a_1 ... a_q\leq N  } J_{a_1...a_q}\psi_{a_1}...\psi_{a_q}, \quad \{\psi_a,\psi_b\}=\delta_{ab}~,
\ee
and restricting to a class of operators whose time evolution can be described by a string of fermions contracted with the $J$ tensor in various ways, they arrive at a very insightful  valid in the large $N$ limit. Namely, they find that the characterization of operator growth can be mapped to the evolution of the wavefunction of a particle performing a quantum walk on a certain kind of graph. The Hamiltonian \footnote{We follow the terminology of \cite{Roberts:2018mnp} which distinguishes between the SYK Hamiltonian $H$ and the Hamiltonian formulation of operator evolution given in terms of $\hat{H}$, which is essentially the Liouvillian.} can then be expressed in terms of its  adjacency matrix as 
\be \label{SYK graph Hamiltonian}
\hat{H}= 2^{1-\frac{q}{2}}J \cdot (\text{adjacency matrix})~,
\ee
where 
\be
\braket{J^2_{a_1...a_q}}=\frac{(q-1)!}{N^{q-1}}J^2~.
\ee

This graph arises by considering all the different ways in which the size of an operator can change. As we will see this is equivalent to identifying the size of an operator with a tree graph. The graph corresponding to the Hamiltonian is then describing all the different tree graphs that can occur through time evolution. More concretely, taking $q=4$ as an example and assuming that the initial operator is of the form $\psi_1(t)$, the authors of \cite{Roberts:2018mnp} make the following identification
\be
\begin{gathered}
    \includegraphics[scale = 2]{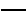}
\end{gathered}
 = \mathcal{O}_0 = 2^\frac{1}{2}\psi_1.
\ee
Namely, the size of the initial operator string can be captured by a single line. By commuting this initial operator with the Hamiltonian, one obtains a string with 3 fermions, which corresponds to
\be
\begin{gathered}
    \includegraphics[scale = 2]{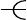} 
\end{gathered}
 = \mathcal{O}_1 = 2^{\frac{3}{2}}\sum_{a<b<c}\frac{J_{1abc}}{J}\psi_a\psi_b\psi_c.
\ee
The interpretation offered is that the original fermion $\psi_1$ has split into $q-1$ fermions by a single commutation with the Hamiltonian. Commuting with the Hamiltonian again, one finds that there are multiple ways of splitting fermions, leading in turn to multiple trees of the following form
\begin{align}
\begin{gathered}
    \includegraphics[scale = 2]{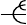}   
\end{gathered}
&= \mathcal{O}_2^{(1)} = 2^\frac{5}{2}\sum_{\substack{a_1<a_2<a_3 \\ b_1<b_2<b_3}}\frac{J_{1a_1a_2a_3}J_{a_1b_1b_2b_3}}{J^2}\psi_{a_2}\psi_{a_3}\psi_{b_1}\psi_{b_2}\psi_{b_3},\label{eq:operator-k-2-ell-1} \\
&\vdots\notag\\
\begin{gathered}
    \includegraphics[scale = 2]{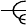} 
\end{gathered}
  &= \mathcal{O}_2^{(3)} = 2^\frac{5}{2}\sum_{\substack{a_1<a_2<a_3 \\ b_1<b_2<b_3}}\frac{J_{1a_1a_2a_3}J_{a_3b_1b_2b_3}}{J^2}\psi_{a_1}\psi_{a_2}\psi_{b_1}\psi_{b_2}\psi_{b_3}.
\end{align}

Generally the growth of an operator can be captured by tree graphs whose vertices have a fixed degree q. We can then consider the graph that catalogs all the different ways an operator can grow, or equivalently all the generations of tree graphs, as the generator of time evolution. Note that this concerns a single instantiation of the model and not an ensemble average, although in the large N limit the SYK model is self-averaging \cite{Rosenhaus:2018dtp}. This is precisely how the SYK Hamiltonian can be brought into the form of (\ref{SYK graph Hamiltonian}). It turns out that this kind of graph has a relatively simple structure related to the Fuss-Catalan numbers, which we will explore in detail below. While, this was already partially investigated in \cite{Roberts:2018mnp}, the relationship with the Fuss-Catalan numbers was not explicit so making this connection here clarifies a lot of the combinatorial properties involved.  

\begin{figure}[ht]
    \centering
    \includegraphics[width=0.7\linewidth]{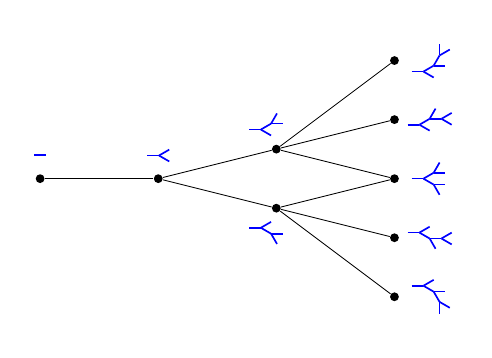}
    \caption{The graph of the first four generations of binary trees. It is well known that the numbers of graphs in subsequent generations are given by the Catalan numbers.}
    \label{binary generations}
\end{figure}

Even though the model has physical significance only for even $q$, we nevertheless find it insightful to start with the simplest non-trivial example $q=3$, which corresponds to the graph of generations of binary trees. In figure \ref{binary generations}
we provide a visualization of the first few generations to give the reader a concrete manifestation of the structures discussed above (see figure 2 of \cite{Roberts:2018mnp}, for the $q=4$ version of the graph).It is well known that the number of binary trees of a given order is given by the Catalan numbers $C_n$. In other words, the number of vertices in each neighborhood is $V_n=C_n$. In order to compute $b_n$, we only need to find the number of edges between adjacent neighborhoods, which we can obtain from a simple counting argument. A binary tree of order $n$ leads to $n$ trees of order $n+1$, so the number of edges between two neighborhoods is simply the product $E_{n-1}=(n+1) C_n$. The Lanczos coefficients are then
\be
b_n= 2^{1-\frac{q}{2}}J\frac{n C_{n-1}}{\sqrt{C_n C_{n-1}}}~.
\ee

More generally, for arbitrary $q$, the number of $k$-ary trees (where $k=q-1$) in a given neighborhood corresponds the (single parameter) Fuss-Catalan number, defined as 
\be
A_k(n)=\frac{1}{k n+1} \binom{k n+1}{k}~,
\ee
which reduces to the usual Catalan numbers for $k=2$. In this case the number of edges between adjacent neighborhoods is $E_{n-1}=2+n(k-1)-k$ and thus the Lanczos coefficients are given by
\be  \label{SYK graph lanczos}
b_n= 2^{1-\frac{q}{2}}J\frac{(2+n(k-1)-k) A_k(n-1)}{\sqrt{A_k(n) A_k(n-1)}}~.
\ee

Let us attempt to retrieve the known expression for the Lanczos coefficients in the large $q$ limit, which is \cite{Parker:2018yvk}
\be \label{SYK lanczos}
b_n=\left\{ 
\begin{array}{cc}
     \mathcal{J} \sqrt{\frac{2}{q}} +O(\frac{1}{q}) \, ,& n=1 \\
   \mathcal{J} \sqrt{n(n-1)} +O(\frac{1}{q}) \, , & n > 1
\end{array}\right. ~,
\ee
with $\mathcal{J}=J\sqrt{\frac{q}{2^{q-1}}}$.
It is evident that we should not expect to get a perfect match. As we saw in section \ref{special bn} the above Lanczos coefficients should correspond to the family of graphs with the characteristic squid shape and that is not the case for the graph we are currently considering. It is straightforward to check by direct substitution in (\ref{SYK graph lanczos}) that  $b_1= \mathcal{J} \sqrt{\frac{2}{q}}$, for any q. For $n>1$ we can expand the Fuss-Catalan numbers as 
\be
A_k(n)=  \frac{n^{n-1}k^{n-1}}{n!}+O(\frac{1}{q})~,
\ee
which leads to 
\be 
b_n=\mathcal{J} \sqrt{n(n-1)}\left(\frac{n-1}{n}\right)^{\frac{n-1}{2}}+O(\frac{1}{q})~.
\ee
This differs from the previous expression, as expected. First, let us point out that it does not differ by a lot in the sense that the extra factor $\left(\frac{n-1}{n}\right)^{\frac{n-1}{2}}$ quickly converges to the constant value $e^{-1/2}$ and thus the scaling of the Lanczos coefficients remains practically the same. In fact for large $n$, the emergent SL(2,R) symmetry is still apparent in line with \cite{Caputa:2021ori}. However, we still have to explain the origin of this discrepancy. In an attempt to do so let us review the derivation of the original result and the differences with our approach. 

In \cite{Parker:2018yvk}, the authors use the infinite temperature, large-$q$ autocorrelation function 
\be
C(t)=1+\frac{2}{q} \ln{\text{sech}(\mathcal{J}t)} + O(1/q^2)~,
\ee
in order to compute the moments of the SYK Hamiltonian. They find that 
\be
\mu_{2n}=\frac{2}{q}\mathcal{J}^{2n}T_{n-1} +O(1/q^2),\quad n>0~,
\ee
where $T_n$ are the tangent numbers. Using the continued fraction expansion of the latter, they obtain a precise formula for the Lanczos coefficients as seen in (\ref{SYK lanczos}). We believe it is most likely that the origin of this discrepancy lies in the use of the large-$q$ autocorrelation function, which seems to effectively coarse-grain the detailed structure of the graph. In other words, starting from this result we would not be able to  derive the expressions for $V_n$ and $E_n$. Our approach on the other hand, relies on the exact tridiagonalization of a single instance of the SYK Hamiltonian, so we are confident that our result is a small, yet important correction to its predecessor. 

Finally, let us reiterate the main points of this section and comment on some of their general implications. We have worked with the SYK model in the large N limit, where the evolution of a single operator can be described as a quantum walk on a graph of the type described in figure \ref{binary generations}. It might be counterintuitive that starting from a Hamiltonian with random interactions the dynamics can be translated to an unbiased walk, but this is explained by the self-averaging property of the model in the large N limit, hence why there is a factor of $J$ in the graph Hamiltonian (Liouvillian) \eqref{SYK graph Hamiltonian}. Implementing our prescription for extracting the Krylov basis from a graph produces an analytic expression for the Lanczos coefficients for any value of $q$. Comparing with the known expression in the large $q$ limit we find a small discrepancy, which we attribute to the previous method of arriving at the Lanczos coefficients, which utilizes a genuine disorder average. This fits into a larger discussion regarding deviations that arise from different ways of averaging, which we elaborate on in section \ref{comparison section}.

More generally, regarding systems with random interactions, we believe a similar approach might be applicable owing to the prediction of ETH that such systems should be self-averaging in the thermodynamic limit. In this way the a priori complicated dynamics can be potentially recast as a quantum walk with fixed edge weight given by the average strength of the interaction.

\subsection{The hypercube} \label{Hypercube section}
The hypercube in $D$ dimensions has the following properties. It has a total of $2^D$ vertices of constant degree $D$. The diameter of the graph is also $D$, which implies that the hypercube can be divided into $D+1$ neighborhoods. The adjacency matrix does not have a standard form in arbitrary dimensions, but it can be defined recursively using the formula
\be
 A_D=
\left\{ 
\begin{array}{cc}
    \mathds{1}_2 \otimes A_{D-1}+A_1\otimes\mathds{1}_{2^{D-1}} \, &, ~D>1 \\
   \begin{pmatrix}
       0 &1\\1&0
   \end{pmatrix}  \,& , ~ D = 1
\end{array}\right. ~.
\ee
It is typical to define the Hamiltonian with the normalization $H_D=\frac{A_D}{D}$, which will also be convenient for the comparison with a classical walk. Technically the results we will be comparing with use a normalization of $D+1$ to account for the option that the  walker stays put, though the approximation $D+1\simeq D$ is then used. If we were to take this into account we would find that we have $a_n=\frac{1}{D+1}$, which however leaves the wavefunctions invariant (up to a phase) and so the expression for Krylov complexity is the same. In order to keep our derivation cleaner, we neglect this effect from the outset and only compute the $b_n$.

We will be using eq. \eqref{graph lanczos} to compute the  Lanczos coefficients of the hypercube graph in an arbitrary number of dimensions. For that we need two pieces of information. The number of vertices in each neighborhood and the total number of edges between two adjacent neighborhoods. The former is given by the binomial coefficient
\be
V_n=\binom{D}{n}~.
\ee
Given that all the vertices have the same degree it is straightforward to show that the total number of edges between two neighborhoods containing $V_{n-1}$ and $V_n$ vertices respectively is $V_{n-1}(D-(n-1))$. Plugging everything in \eqref{graph lanczos} we obtain 
\be
b_n=\frac{1}{\sqrt{V_n V_{n-1}}}\frac{V_{n-1}(D-(n-1))}{D}=\frac{1}{D}\sqrt{n(D-n+1)}~.
\ee

Note that this is the exact same formula for the Lanczos coefficients of a system with SU(2) symmetry \cite{Caputa:2021sib} with $2j=D$, being the quantum number associated with the particular representation of the group. It is straightforward to argue that the neighborhood states of the hypercube furnish a representation of SU(2). This stems from the fact that the dimension of their space is, as explained by the simple counting argument above, $D+1=2j+1$. Since the dimension of this vector space is equal to the dimension of a representation of SU(2) with weight $j$, it follows that the two spaces are isomorphic.
We can further refine this argument by utilizing the ``complexity algebra" introduced in \cite{Caputa:2021sib}. In short, what the latter asserts is that since the Hamiltonian is of the form (\ref{neighborhood Hamiltonian}), it can be understood as being comprised by a pair of ladder operators, hence 
\be
H=\alpha (J^\dagger+J)~.
\ee
This further implies the existence of an operator $B=\alpha(J^\dagger-J)$ and the product of its commutator with $H$, which is diagonal in the Krylov basis, thus denoted by $\tilde{K}=-4\alpha J_0$. It has been shown that for a system with SU(2) symmetry, this algebra is closed with the condition
\be
2(b_{n+1}^2 -b_n^2)=An+B~,
\ee
with $A,B$ being some constants. Using the $b_n$ we computed above, we find that this equation is satisfied with $A=-\frac{4}{D^2}$ and $B=\frac{2}{D}$. This proves the closure of the complexity algebra. Furthermore, it is easy to verify that the eigenvalues of $\tilde{K}$ are of the form $\tilde{k}=-\frac{4}{D^2}(n-j)$ which proves that it is in fact the algebra of the group SU(2) in accordance with the results of \cite{Caputa:2021sib}.
It follows that the wavefunctions and Krylov complexity for a quantum walk on the $D$-dimensional hypercube are given by 
\begin{align}\label{SU(2) wf}
&\varphi_n(t)=\frac{\tan{\left(\frac{t}{D}\right)}^n}{\cos{\left(\frac{t}{D}\right)}^{-D}}\sqrt{\frac{\Gamma(D+1)}{n! \Gamma(D-n+1)}}\\
&\mathcal{C}_K(t)= D \sin{\left(\frac{t}{D}\right)}^2~. \label{hypercube complexity}
\end{align}
Despite this being a well-known result, we will find that the graph perspective has important implications about our general understanding of complexity.

\section{Comparison with the complexity of random unitary circuits } \label{comparison section}
The example of the hypercube offers an opportunity to compare our results regarding Krylov complexity with previously derived results for the circuit complexity arising from a classical random walk, as presented in \cite{Gautason:2025ryg}. There, the authors use a prescription for the computation of complexity by studying discrete subgroups of the unitary group, first given in \cite{Lin:2018cbk}. Such discretizations reduce the geometry of the space of unitaries to that of a Cayley graph, an example of which is the hypercube itself.  Therefore, a trajectory on the graph is nothing but a discretized version of a continuous trajectory in the space of unitaries. Implementing a classical random walk allows one to efficiently explore the aspects of this discrete geometry, but more importantly it provides a clear picture of this process in terms of a quantum circuit built out of a string of unitaries chosen at each time step of the walk. 

More concretely, this picture arises by considering the implementation of two-qubit gates acting on all qubits at every time step of the circuit. A graph can then be used to visualize this process, wherein every vertex at a given distance from the origin represents one possible combination of such pairings between qubits. If one wishes to study a random circuit (meaning at every time step the gates are chosen randomly), this is equivalent to a random walk on the associated graph. The depth of the random circuit can be thus identified with the expected distance of the walker from its initial position. Note that the randomness is externally imposed rather than inherent, which is a point we will return to.

While this framework provides a concrete manifestation of the use of classical random walks to study circuit complexity, the core idea goes back to the early works on the topic \cite{Susskind:2014jwa,Brown:2016wib,Brown:2015bva,Brown:2015lvg,Lin:2018cbk,Susskind:2018pmk}, which use the concept of a classical random walk in the space of unitaries as a conceptual vehicle leading to the characterization of qualitative features of complexity in holographic systems. A lot of the assumptions therein rest on the features of the Hayden-Preskill circuit \cite{Hayden:2007cs}. We will progressively address each one of these issues below. 

\subsection{Two approaches for the hypercube}

In \cite{Gautason:2025ryg} the authors use the average distance from the origin of the hypercube as a measure of complexity, exactly like we did in the previous section. However, due to the classical nature of the walk, they find as their definition for complexity
\be \label{circuit hypercube}
C(t)\equiv\braket{x(\tau)}=\frac{D}{2}(1-e^{-2\tau/D})~,
\ee
where $x(\tau)$ denotes the distance as a function of the time $\tau$ counting the time steps. In other words, circuit complexity is identified with the expected distance from the initial vertex as a function of time. By comparing with the Krylov complexity associated to a \emph{quantum} walk on the same hypercube (\ref{hypercube complexity}) it is clear that the two versions of complexity differ dramatically, at least when taken at face value. This is not entirely unexpected given that we are comparing two different measures, so one might have predicted a quantitative difference between the two.  
For this reason, several comments are in order.

First, the mathematical origin behind this discrepancy is clear. A classical walker on the hypercube is likely to move towards the middle of the graph quite quickly, but then the probability of moving either to the left or to the right is equal and as such they will spend most of their time around that neighborhood. We can easily observe that behaviour by considering the early and late time limits of (\ref{circuit hypercube}) for which complexity grows as $C(\tau)=\tau + O(\tau^2)$  and saturates at the value $C(\tau)=\frac{D}{2}$ respectively. This is precisely the feature that the authors of \cite{Gautason:2025ryg} utilize in order to draw parallels with the dynamics of a black hole, which is expected to exhibit this initial linear growth, followed by saturation at times of order $t_{sat}=\frac{1}{4}e^S$. In their setup this holds true if one identifies $D\sim e^{S}$, with $S$ being the entropy of a black hole. On the contrary, a quantum walker's wave function spreads uniformly over the whole graph and the average distance will oscillate in time, exhibiting neither linear growth, nor saturation. 

The question that naturally arises then is why are we led to these two different pictures? The short answer is averaging. We will relegate a more thorough explanation to section \ref{HP and avg}, however we can easily see that there is a simple way to rectify, at least in part, the discrepancy between these two approaches if we consider the time average of Krylov complexity, defined in (\ref{time average complexity}). For the hypercube this leads to the expression
\be\label{eq:avgcube}
\overline{\mathcal{C}_K(T)}=\frac{D}{2} \left(1-\frac{D}{2 T} \sin \left(\frac{2 T}{D}\right)\right)~,
\ee
which for early times grows as $\overline{\mathcal{C}_K(T)}=\frac{T^2}{3} +O(T^3)$ and at late times saturates at precisely $\overline{\mathcal{C}_K(T)}=\frac{D}{2}$. We thus find the same trend of growth and saturation, though we should be careful in extracting a physical interpretation from it, given that this appears to be a generic feature of systems with a finite number of degrees of freedom (in this case finite graphs). Moreover, we can show that the time that it takes to reach the saturation value is faster for Krylov complexity. For the classical walk the saturation can be estimated to happen at the time the average distance is one standard deviation away from the saturation value, i.e.
\be
\braket{x(\tau_{\text{sat}})}=\frac{D}{2}-\frac{\sqrt{D}}{2}~,
\ee
which leads to $\tau_\text{sat}=\frac{1}{4}D \log D$. Within our approach, the standard deviation is not a constant, so estimating the saturation time is not as straightforward. We can start by computing the standard deviation using the wavefunctions in (\ref{SU(2) wf})
\be
\sigma_K(t) =\left(\sum_{n=0}^D n^2 \abs{\varphi_n}^2-\left(\sum_{n=0}^D n \abs{\varphi_n}^2\right)^2\right)^{\frac{1}{2}}=\frac{D}{2}\sin{\left(\frac{2t}{D}\right)}~.
\ee
Subsequently we can compute the moving average 
\be
\overline{\sigma_K(T)}= \frac{1}{T}\int_0^T \sigma_K(t)dt= \frac{D^2}{2T}\sin{\left(\frac{T}{D}\right)^2}~.
\ee
A reasonable choice is to define the saturation time as the first zero of $\overline{\sigma_K(T)}$, since at the point the complexity will be exactly $\overline{C_K}=\frac{D}{2}$ and subsequently it will fluctuate around that value with ever smaller fluctuations. 
We will see that this choice is of little consequence, especially when the hypercube dimension becomes large. The solution we obtain is 
\be
\overline{\sigma_K(T_\text{sat})}=0
\quad
\Rightarrow
\quad T_\text{sat}=\pi D~,
\ee
which evidently scales more slowly compared to $D\log D$. It should now be apparent that the choice of the exact saturation time for our quantum walk is not hugely important, as the scaling will always be linear in $D$. Note that this is the well documented difference between the mixing times of a classical and a quantum random walk on the hypercube \cite{Moore:2001zik}.
\begin{figure}[ht]
   \begin{subfigure}{0.5\textwidth}
       \includegraphics[width=0.9\linewidth]{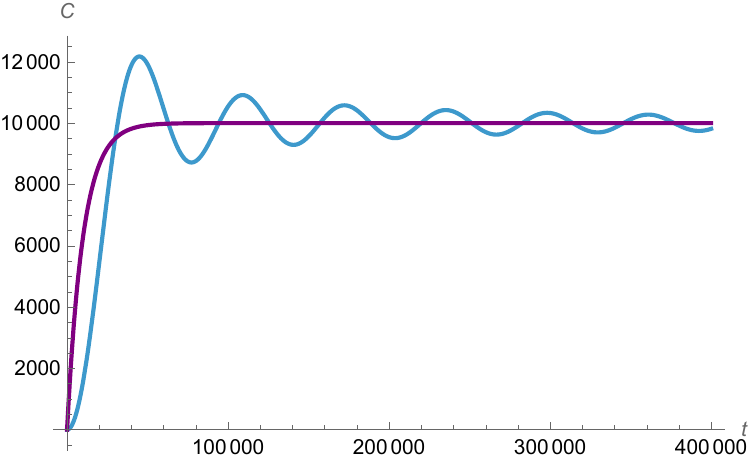}
       \caption{}
\label{av dist}
  \end{subfigure}
\begin{subfigure}{0.5\textwidth}
    \includegraphics[width=0.87\textwidth]{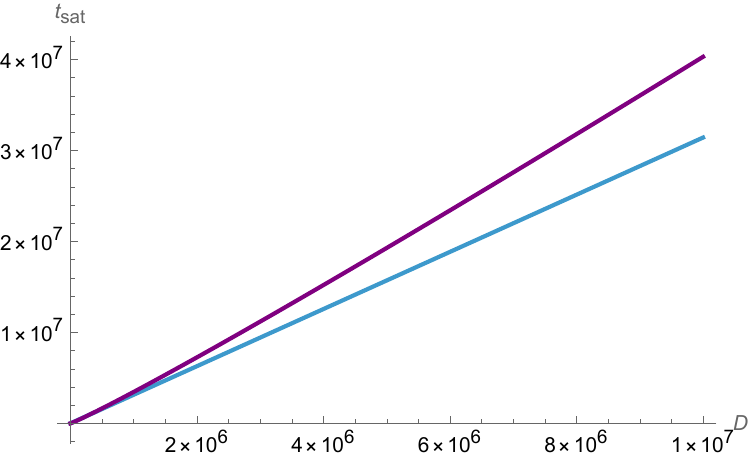}
    \caption{}
\label{tsat}
\end{subfigure}
\caption{(a) Comparison of the time average of Krylov complexity (blue) and circuit complexity (purple) as functions of time on the hypercube with $D$=20.000 (b) Comparison of the saturation time for a quantum (blue) and classical (purple) random walks as a function of the hypercube dimension D.}
\label{ruc vs krylov}
\end{figure}
In figure \ref{ruc vs krylov} we present a comparison of the complexity and saturation times between the classical and quantum walks on the hypercube. It appears that in spite of the different methods used, the picture we obtain after averaging is quite similar, though the fact remains that for very large $D$, Krylov complexity is expected to saturate earlier. 
We note that the characteristic time scale $\pi D$ also arises as the oscillation time period of the averaged Krylov complexity \eqref{eq:avgcube}.

\subsection{Walks in the space of unitaries}

Having discussed this concrete example, we can go back to the more conceptual picture describing the exploration of the space of unitaries $SU(2^K)$ by means of a random walk on a graph and its circuit equivalent, as offered in \cite{Susskind:2018pmk}. The presupposition in this case is that the structure of the graph is that of a $q$-regular decision tree. This is justified on the basis of possible pairings of qubits in the auxiliary circuit whose complexity we wish to study, which however does not explain the classical nature of the walk. After all we are studying the evolution of a quantum system and by asserting that we are simply pairing the qubits at random does not provide any clear transition from quantum to classical. So where did all the ``quantumness" go? We will examine this point closely in the next section, but for now let us just accept it in good faith and see to what conclusions it can lead us. 

Notice that the graph under consideration precisely belongs in the family given by (\ref{exp graps}) for $c_1=c_2=c_3=1$. It is quite straightforward then to compare how Krylov complexity behaves in comparison with the established circuit picture. The Lanczos coefficients are simply 
\be
b_n=\sqrt{q}~.
\ee
For an infinite Krylov space this would imply linear growth of complexity, matching the intuition from the circuit picture. In reality though we are looking to describe systems with finite size, since black holes have a finite, albeit typically very large, entropy. Hence, when we take the finite size effects into account we will see that once again the picture changes dramatically. Let us then treat this problem in more detail using the Lanczos coefficients above.

\begin{figure}[ht]
    \centering
    \includegraphics[width=0.5\linewidth]{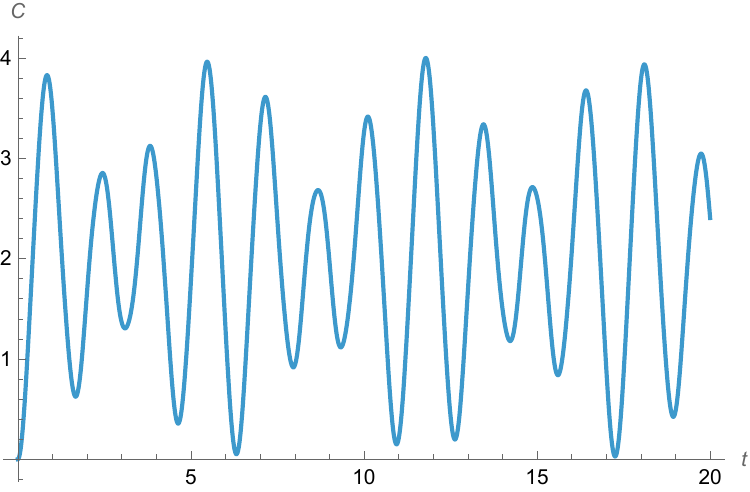}
    \caption{Krylov complexity as a function of time for $q=16$ and $N=5$.}
    \label{constant bn complexity}
\end{figure}

The fact that they are constant implies that the Hamiltonian is a Toeplitz matrix, for which there exists a closed expression for the eigenvalues (not to be confused with the expression for the number of edges between adjacent layers)
\be
E_j=2\sqrt{q} \cos{\left(\frac{\pi j }{N+1}\right)}~.
\ee
Where $j$ runs up to the total number of eigenvalues $N$, which in turn is related to the number of sites on the Krylov chain.
Since we have this information at our disposal we can use it to obtain the wavefunctions $\varphi_n$, rather than building them using the survival amplitude given by $\varphi_0$ and the Lanczos coefficients themselves. Following the recipe used in \cite{Balasubramanian:2025xkj} for a closely related class of problems involving the tight-binding model on a chain, we get 
\be
\varphi_n(t)=\frac{2}{N+1}\sum_{j=1}^N \sin{\left(\frac{\pi (n+1)j}{N+1}\right)}  \sin{\left(\frac{\pi j}{N+1}\right)} e^{- 2 i\sqrt{q} t \cos{\left(\frac{\pi j}{N+1}\right)}}~.
\ee
From this expression, it is straightforward to compute Krylov complexity for any finite $N$ and $q$. Unsurprisingly we find that the complexity oscillates around the value $\frac{N-1}{2}$ in agreement with \cite{Balasubramanian:2025xkj}, as can be seen in figure \ref{constant bn complexity}.

Upon averaging we find that the same pattern of growth and saturation arises. Even though we do not have a closed formula for the complexity at our disposal, we can show that it saturates at the value $\overline{C_K(T)}= \frac{N-1}{2}$ as follows
\begin{equation}\begin{aligned}
	\lim_{T\to\infty}\frac{1}{T} \int^{T}_{0}dt ~\mathcal{C}_{K}(t)
	=&~
	\sum_{n,j,j'}n
	\langle \psi(0)|E_{j}\rangle\langle E_{j}|n\rangle
	\langle n|E_{j'}\rangle\langle E_{j'}|\psi(0)\rangle
	\lim_{T\to\infty}\frac{1}{T} \int^{T}_{0}dt e^{-i(E_{j}-E_{j'})t}
    \\
    =&~
	\sum_{j}|c_{j}|^{2}
	\mathcal{O}_{jj}~,
\end{aligned}\end{equation}
where $c_j=\langle E_{j}|\psi(0)\rangle$ and
\begin{equation}
	\mathcal{O}_{jj}
	=
	\langle E_{j}|
	\hat{K}	
	|E_{j}\rangle 
	, 
	\quad \hat{K}=\sum_n n\ket{n}\bra{n}~.
\end{equation}
Notice that the above formulation looks suspiciously similar to the eigenstate thermalization hypothesis (ETH), a point on which we comment in the following section. Given the above we can calculate
\be
\Op_{jj}= \frac{N-1}{2}~. 
\ee
Since this is a constant and $\sum_j\abs{c_j}^2=1$, it follows that 
\be
\lim_{T\to\infty}\frac{1}{T} \int^{T}_{0}dt ~\mathcal{C}_{K}(t)
	= \frac{N-1}{2}~.
\ee

Calculating the timescale of saturation is more challenging, but at first glance the comparison with the classical random walk approach appears to yield a similar picture. In order to properly account for the finite size effects of the black hole \cite{Susskind:2018pmk} suggests that we consider expander graphs. Remarkably, these graphs are known not to exhibit speed-ups in terms of the mixing time \cite{Aharonov:2002cmr}. This means that we can expect both Krylov and circuit complexities to saturate at the same timescale.

\subsection{Hayden-Preskill and the need for averaging} \label{HP and avg}
With the above considerations and results in mind we are ready to investigate the heart of the issue, namely the complexity as it arises through Hamiltonian evolution as opposed to a random unitary circuit of the Hayden-Preskill type. Let us point out that the idea of simplifying unitary dynamics to a circuit, which is itself efficiently described by a classical stochastic process has been the subject of many seminal works apart from Hayden and Preskill's, such as \cite{Znidaric:2008idn,Dahlsten:2007uow,Sekino:2008he,Lashkari:2011yi,Khemani:2017nda,Nahum:2017yvy}. Some of the main ideas therein are succinctly summarized in \cite{Fisher:2022qey}.

We find the explanation presented in \cite{Nahum:2017yvy} most illuminating. The authors consider an operator evolved by a unitary circuit $O(t)=U^\dagger(t)OU(t)$, which can be expanded in the basis of $SU(q^N)$ ($q$ here stands for the dimension of the Hilbert space at each of the $N$ sites) generators $\{\mathcal{S}\}$ as 
\be
O(t)=\sum_\mathcal{S}a_\mathcal{S}(t)\mathcal{S}~.
\ee
They then find that the probability  $P_\mathcal{S}(t)\equiv\overline{a^2_\mathcal{S}(t)}$ satisfies the master equation for a Markov process
\be
P_\mathcal{S}(t)=\sum_{\mathcal{S}'} W_{\mathcal{S}\mathcal{S}'}P_{\mathcal{S}'}(t-1)~,
\ee
where $W_{\mathcal{S}\mathcal{S}'}$ is a suitable transition matrix whose properties are not important for our purposes here. Notice however, that the probability $P_\mathcal{S}(t)$ is quite crucially defined as the Haar average $\overline{a^2_\mathcal{S}(t)}$. However, as the authors caution, ``this fictitious Markov process is not the true unitary dynamics of $O(t)$". This clarifies why we should not expect a match between a measure of complexity derived from this Markov process and one derived directly from the unitary dynamics.

While this stochastic process above is not precisely identified with the classical walk of the previous subsections, it is a clearer manifestation of the implicit coarse-graining procedure that leads to the latter. More concretely, the fictitious ensemble of walkers in \cite{Susskind:2018pmk}, is what leads under the no-collisions assumption (equivalently assuming the graph is tree-like) to the circuit being Haar typical. In other words, no matter the starting point and whether you assume that the gates are chosen randomly with respect to the Haar measure as in \cite{Nahum:2017yvy}, or the gates are fixed but their implementation is random as in \cite{Susskind:2018pmk}, averaging is imperative for classical randomness to emerge.

Be that as it may, this picture captures many of the essential features of operator growth, which is precisely why it has been so influential. The common points with our approach seem to arise when we consider the time average of Krylov complexity (\ref{time average complexity}). In light of the above, we can provide an explanation for this phenomenon. The systems we are considering exhibit a version of ergodicity, since by construction the time evolution leads to a complete exploration of the space of states. Simply put, since we are looking at connected graphs, the wavefunction is bound to eventually spread to all the vertices. It is then reasonable to assume that performing the time average has a similar effect to an ensemble average. This is precisely why we can cast the problem at hand in the language of ETH.

Summarizing the main points of this section, we have shown that the Krylov complexity for a quantum walk on a finite graph will generically exhibit oscillatory behaviour, rather than growth and saturation. Upon averaging, we obtain partial agreement with the expected picture, with the difference that the saturation time potentially experiences a speed-up and differing scrambling time behaviour, depending on the graph at hand. The fact that saturation can happen earlier than expected is puzzling. 
The very point of utilizing complexity in this context is to describe the system after it has thermalized, so it is important that the former keeps growing for a long time. 
\section{Discussion and outlook} \label{discussion}
Let us recapitulate the main results of this work and comment on their implications and possible extensions. In this work we have provided a framework for the computation of the Lanczos coefficients and Krylov complexity for a quantum walk on an arbitrary graph. Conversely, we have shown that we can assign a family of graphs to different classes of quantum systems with well-known complexity properties. Additionally, we have used different examples to highlight the utility of our approach. In particular, we implemented the methods we developed on a graph that describes the operator growth in the SYK model and have obtained a simple expression for the Lanczos coefficients for an arbitrary number of interacting fermions $q$. We also extensively studied the hypercube and used it as a springboard to compare our results with the circuit complexity arising from a classical walk on the same graph. 
We found that the late time Krylov complexity, which is purely oscillating, can only be reconciled with the classical result, which has a ubiquitous converging behaviour. To arrive at this result, we had to perform time averaging on the Krylov complexity, which we can relate to the classical result through an ETH type of reasoning. Some differences, even after averaging, remain. Namely, the early time behaviour is linear in time for the classical result, whereas we obtain a quadratic in time behaviour from Krylov complexity. Note that this is a universal feature of Krylov complexity so this is precisely what one would expect, even though the reason for the difference between the two approaches remains unclear. Lastly, we report a speed up of convergence in the Krylov case as compared to the classical case.

From this point there are many paths one can take to expand beyond the work we present. For example, we mentioned that our prescription should also be applicable to biased quantum walks, where the weights of the edges of the graph are different. 
Amongst others this would open the doors to studying open quantum systems, inherently related to non-Hermitian Hamiltonians.
One problem of particular interest which is offered for investigation concerns operator growth in 2D CFTs.  In \cite{Caputa:2021ori} the authors study the growth of primary operators and an important aspect of their construction is the Young lattice describing the Verma module structure of a given primary and its descendants. In essence, the problem of describing operator growth reduces to a biased quantum walk on the Young lattice, where the transition probabilities between adjacent sites are given by the rules of CFT. The Young lattice treated as a graph has a well defined structure, meaning that there exists a closed expression for the number of vertices in a given neighborhood, simply given by the number of integer partitions $V_n=p(n)$. From that information one can derive the number of edges to be the cumulative sum of integer partitions 
\be
E_{n-1}= \sum_{n=0}^Np(n)~.
\ee
In principle one can apply (\ref{graph lanczos}) to compute the Lanczos coefficients, as long as the appropriate weights for the edges are accounted for. Given that the solution of this problem was already partially given in \cite{Caputa:2021ori} it would be interesting to see what our methods yield and whether they match their results. While we believe this to be a tractable problem, it is nevertheless challenging enough so as to remain beyond the scope of this initial exposition of the topic. 

Another potentially fruitful direction concerns further investigation on the matter of holographic complexity. As we mentioned in the introduction much effort has been devoted to the study of Nielsen complexity, which is a more refined measure compared to circuit complexity as it inherently characterizes Hamiltonian evolution. While its relationship with Krylov complexity has been thoroughly studied in a few works \cite{Lv:2023jbv,Craps:2023ivc,Craps:2025kub}, it would be interesting to see how these results come together in the context provided here, especially in light of the recent results of \cite{Craps:2025kub} regarding the relationship of Krylov and Nielsen complexities in the context of the SYK model. 

Within this same class of problems, we have detected yet another potential connection of our work to previous results regarding the SYK model and in particular, its double scaling limit. It has now been well understood that within the context of the double scaled SYK (DSSYK), the Krylov basis can be identified with the so-called chord basis \cite{Rabinovici:2023yex,Ambrosini:2024sre,Lin:2022rbf,Berkooz:2023cqc,Berkooz:2023scv}. The chord diagrams from which this basis takes its name are used to extract the moments of the Hamiltonian. Incidentally, chord diagrams share similar combinatorial properties with tree graphs and the related Fuss-Catalan numbers we used to describe them, so it is possible that there exists an overarching framework for these two seemingly related problems. 

Our findings have implications for black hole physics when invoking black hole complementarity: the view that an outside low-energy observer can interpret any physical processes associated to the black hole by thinking of them as a collection of qubits governed by some Hamiltonian. When treating these qubits using an ETH-inspired averaging, we find the ramp-plateau type structure that is also reproduced through gravitational computations -- for example the growth of the volume of wormhole. An open question is: what is the quantity that returns the non-averaged Krylov complexity results? Or should one even expect the geometry to have access to this information? And if so, how can one probe it from the geometry?

Finally, we should not fail to mention the prospective applications geared towards quantum algorithms and their potential advantages over classical ones. Generally, finding quantum algorithms exhibiting speed-ups is one of the necessary ingredients for quantum computing. However, the list of concrete examples is rather limited. Here we provided a framework which can prove useful in detecting and constructing such examples, as well as characterizing them using the fundamental tools of quantum mechanics. While quantum walks have been investigated in the same spirit for example in \cite{Balasubramanian:2023qoi} and Krylov complexity has garnered attention in relation to quantum computing (see for instance \cite{Cindrak:2024htv,Cindrak:2025jfj}), we believe the framework we provided makes for a more robust bridge between these topics and an arena in which these ideas can be tested more thoroughly.

In conclusion, the interface between quantum walks and the study of complexity as it has been shaped by the intense focus of the high energy community is rife with questions touching upon the rudiments of a variety of cutting edge topics. As such, we are convinced that further exploration of this intriguing connection is  a promising new direction and we are looking forward to novel and exciting results.

\section*{Acknowledgments} 
We would like to thank P. Caputa and G. di Giulio for useful discussions and early feedback. We also thank B. Chen, F. Gautason, V. Mohan, T. Schuhmann and L. Thorlacius for comments on the draft. 

DP wishes to especially thank WS and Nordita for their gracious hospitality during the initial and final stages of this project, as well as M. Panfil and M. Lisicki who through their course on stochastic processes inspired many of the ideas in this work. The work of WS is supported by Starting Grant 2023-03373 from the Swedish Research Council and the Olle Engkvist's Foundation. 
Both authors are grateful to COST Action CA22113, ``Fundamental challenges in theoretical physics" for making this collaboration possible.

\appendix

\section{Graph symmetry and equitable partitions} \label{graph conditions}

The aim of this appendix is to address the conditions under which the construction of the neighborhood states (\ref{neighborhood states}) leads to the Krylov basis. The first ingredient we need is the notion of an equitable partition, which is the partition of the vertex set of a graph such that every vertex in a given part has the same number of neighbors in a different part. More formally, for a graph $G$, a partition $C_1,C_2,...,C_r$ is equitable if the number of neighbors in $C_j$ of a vertex $\alpha$ in $C_i$ is a constant $d_{ij}$ independent of $\alpha$ \cite{godsil2013algebraic}. In other words every vertex in a given part has the same number of neighbors belonging to any other part.

The definition of the so-called column states of \cite{Childs:2001xhf} as well as our generalized notion of neighborhood states relies upon the existence of such an equitable partition in terms of the distance from an initial vertex or set of vertices. This is identified with the notion of the neighborhood partition defined in the main text. This structure in turn guarantees that these states span an invariant subspace under the action of the adjacency matrix, which for our purposes is also the Hamiltonian. Let us check this explicitly. Given a graph $G$ and assuming there exists a distance equitable partition $C_1,C_2,...,C_r$, the definition of neighborhood (column) states is simply
\be \label{partition states}
\ket{c_i}=\frac{1}{\sqrt{\abs{C_i}}}\sum_{\alpha\in C_i}\ket{\alpha}~.
\ee
The action of the adjacency matrix then follows
\be
A \ket{c_i}=\sum_j d_{ij} \sqrt{\frac{\abs{C_j}}{\abs{C_i}}}\ket{c_j}~.
\ee
Since we are restricting to a partition in terms of distance, all the neighbors of $\alpha \in C_i$ are in $C_{i\pm1}$, which in turn lets us rewrite the above formula as 
\be
A\ket{c_i}=d_{i,i+1} \sqrt{\frac{\abs{C_{i+1}}}{\abs{C_i}}}\ket{c_{i+1}}+ d_{i,i-1}\sqrt{\frac{\abs{C_{i-1}}}{\abs{C_i}}}\ket{c_{i-1}}~.
\ee
This shows that the subspace spanned by the neighborhood states is invariant under the action of the adjacency matrix (Hamiltonian), or in other words neighborhood states are only mapped to neighborhood states under the action of $A$. Furthermore, the distance partition forces $A$ to be tridiagonal in that basis. 

The second and last ingredient we require in order to show that this uniquely determines the Lanczos coefficients and Krylov basis is the Lanczos uniqueness theorem \cite{Viswanath1994TheRM,Parlett} which states that given a general operator $\Op$ and a basis of orthogonal vectors $Q=(q_1,q_2,...q_n)$, the tridiagonal form of A given by $Q^*\Op Q=T$ is uniquely determined by $\Op$ itself and $q_1$ or any $q_n$. In other words if one is able to find a tridiagonal representation of the Hamiltonian given an initial state, this guarantees that the matrix elements are the Lanczos coefficients and the associated orthonormal basis is the Krylov basis. This is exactly what we have achieved for the adjacency matrix of a graph by using distance equitable partitions, whence it follows that the neighborhood states uniquely correspond to the Krylov states. 

Finally, let us comment on the concrete implications for the types of quantum walks and graphs we can address using the framework above. As we discussed above the only condition is the existence of a distance equitable partition. For regular graphs (every vertex has the same number of neighbors), it is easy to verify that such a partition exists regardless of the choice of initial vertex or vertices. However, we have also considered examples of irregular graphs for which such a partition naturally arises, as is the case for the graph of operator evolution in the SYK model. In those cases the guiding principle for choosing this particular partition of a graph is the underlying physical assumptions, such as the notion that the initial state or operator is ``simple" and is gradually becoming more complex due to the dynamics. We should caution though that in these cases an equitable partition with respect to a different vertex/vertices might not exist and our prescription would not apply directly. Altering it to fit this scenario is possible, but it is simply as difficult as performing the Lanczos algorithm itself. A notable exception is the case when the weights of edges between individual neighborhoods are equal in which case the Lanczos coefficients are simply multiplied by a constant. More generally though, even for graphs that do not strictly admit a distance equitable partition, they can be treated similarly as long as they do so even approximately. This is for example what underpins the random graph analysis of \cite{Balasubramanian:2023qoi}.

\bibliographystyle{JHEP}
\bibliography{references.bib}

\end{document}